# The mixed convolved action

Gary F. Dargush, Jinkyu Kim

*Department of Mechanical and Aerospace Engineering*
*University at Buffalo, State University of New York*
*Buffalo, NY 14260 USA*

gdargush@buffalo.edu, jk295@buffalo.edu

**Abstract**

A series of stationary principles are developed for dynamical systems by formulating the concept of mixed convolved action, which is written in terms of mixed variables, using temporal convolutions and fractional derivatives. Dynamical systems with discrete and continuous spatial representations are considered as initial applications. In each case, a single scalar functional provides the governing differential equations, along with all the pertinent initial and boundary conditions, as the Euler-Lagrange equations emanating from the stationarity of this mixed convolved action. Both conservative and non-conservative processes can be considered within a common framework, thus resolving a long-standing limitation of variational approaches for dynamical systems. Several results in fractional calculus also are developed.

## 1. Introduction

In the first half of the nineteenth century, Hamilton (1834, 1835) formulated a general method for dynamics based upon the concept of stationary action, with action represented as the integration over time of the Lagrangian function of the system. For systems of particles, the stationarity of this action provides the equations of motion as its Euler-Lagrange equations. Hamilton's principle is, of course, one of the pillars of classical dynamics and has broad applicability in mathematical physics and engineering. There are, however, two main difficulties. The first involves the requirement that the variations at the beginning and end of the time interval are zero. This implies that the positions of the particles are known at those two instants. If we were dealing with a boundary value problem having essential conditions specified at both ends of the domain of interest, then this would be perfectly logical. However, for initial value problems, how can one assume that the position of each particle at the end of the time interval is known, when in actuality the primary objective of such problems in classical dynamics is to find those positions? The second difficulty is the restriction to conservative systems. Although many attempts have been made to alleviate this shortcoming, for example, by introducing a dissipation function (Rayleigh, 1877), the development of a single scalar Lagrangian-based action functional for non-conservative systems has not been possible, unless artificial variables and equations are also introduced.

In Section 2, we review the classical variational approach of Hamilton for dynamical systems and several of the more prominent attempts that have been proposed to extend the formulation to include dissipation. We use the single degree of freedom oscillator as a canonical example. Then, in Section 3, we highlight the work by Gurtin and by Tonti that can address dissipative systems and the issue of proper end point conditions. As we shall see, these efforts appear to be in the right direction, but several more ideas are needed to fully resolve the issues. The first of these originates from fractional calculus and several relevant definitions and results are presented in Section 4. The final piece comes by viewing the system from a mixed variational framework, which is then developed in Section 5 for the damped single degree of freedom oscillator. The mixed convolved action presented there recovers the governing differential equation of motion for this dissipative system, along with the specified initial conditions, as its Euler-Lagrange equations, while allowing free variations at the end of the time interval. The approach is quite general and is subsequently extended to multi-degree of freedom dynamical systems and to linear elastodynamic continua in Sections 6 and 7, respectively. In these two sections, the approach is not always given in full generality, but rather as a means to illustrate potential applicability. A few conclusions are then provided in Section 8.

**2. Classical approaches for single degree-of-freedom oscillators**

*2.1. Hamilton's principle*

Consider the forced single degree-of-freedom dynamic system displayed in Figure 1, consisting of a mass $m$ and linear spring having constant stiffness $k$, under the influence of a known applied force $\hat{f}(t)$. Let $u(t)$ represent the displacement of the mass from its equilibrium position, while

$$v(t) = \dot{u}(t) \tag{1}$$

denotes its velocity with the superposed dot indicating a derivative with respect to the argument, which in this case is the time $t$.

The initial value problem associated with this system consists of the differential equation of motion

$$m\ddot{u} + ku = \hat{f} \tag{2}$$

along with the initial conditions

$$u(0) = u_0; \quad \dot{u}(0) = v_0 \tag{3a,b}$$

Based upon a classical analytical dynamics approach (*e.g.*, Goldstein, 1950; José and Saletan, 1998), we can define the Lagrangian $L$ for this system as

$$L(u,\dot{u};t) = T(\dot{u};t) - U(u;t) + V(u;t) \tag{4}$$

with kinetic energy

$$T(\dot{u};t) = \tfrac{1}{2} m \left[\dot{u}(t)\right]^2 \tag{5a}$$

elastic strain energy

$$U(u;t) = \tfrac{1}{2} k \left[u(t)\right]^2 \tag{5b}$$

and

$$V(u;t) = \hat{f}(t) u(t) \tag{5c}$$

representing the potential of the applied loads.

The action functional $I$ for the fixed time interval from $0$ to $t$ is written

$$I(u,\dot{u};t) = \int_0^t L(u,\dot{u};\tau) d\tau \tag{6}$$

Notice from (4), (5) and (6) that the initial conditions (3a,b) play no role in the definition of the Lagrangian or the action.

For stationary action, the first variation of (6) must be zero. Thus,

$$\delta I = \delta \int_0^t L(u, \dot{u}; \tau) d\tau = 0 \tag{7}$$

or

$$\delta I = \int_0^t \left[ \frac{\partial L}{\partial \dot{u}} \delta \dot{u} + \frac{\partial L}{\partial u} \delta u \right] d\tau = 0 \tag{8}$$

and finally

$$\delta I = \int_0^t \left[ m\dot{u}\delta\dot{u} - ku\delta u + \hat{f}\delta u \right] d\tau = 0 \tag{9}$$

After applying integration by parts to the first term within the integral in (9), this becomes

$$\delta I = -\int_0^t \left[ m\ddot{u} + ku - \hat{f} \right] \delta u \, d\tau + \left[ m\dot{u}\delta u \right]_0^t = 0 \tag{10}$$

Following Hamilton (1834, 1835), in order to recover the governing equation of motion (2), we must invoke the condition of zero variation at the beginning $\delta u(0)$ and end $\delta u(t)$ of the time interval. Then, (10) reduces to

$$\delta I = -\int_0^t \left[ m\ddot{u} + ku - \hat{f} \right] \delta u \, d\tau = 0 \tag{11}$$

Finally, after allowing arbitrary variations $\delta u$ between the endpoints, we have the equation of motion for forced vibration as the Euler-Lagrange equation associated with the stationarity of the action $I$. Thus, (2) is recovered from (11) for arbitrary $\delta u$ at each instant of time $\tau$ with $0 < \tau < t$. Of course, we also can arrive at this equation of motion by invoking the Lagrange formalism and writing

$$\frac{d}{dt}\frac{\partial L}{\partial \dot{u}} - \frac{\partial L}{\partial u} = 0 \tag{12}$$

Either way, nothing can be said concerning the initial conditions, which must be enforced separately, outside of Hamilton's principle.

Taking the second variation of $I$ starting from (9), we find

$$\delta^{(2)} I = \int_0^t \left[ m(\delta\dot{u})^2 - k(\delta u)^2 \right] d\tau \tag{13}$$

Since the arbitrary variations $\delta u$ and $\delta \dot{u}$ are independent, $\delta^{(2)}I$ is indefinite. Consequently, Hamilton's principle for this spring-mass system is associated with stationary action.

In the absence of the spring, however, the response of the mass due to a force derivable from a potential is associated with a minimum of the action functional, as can be seen from (13) with $k = 0$. Then, of course, we have a least action principle.

While a wide variety of formulations based on Hamilton's principle have had a significant impact on mechanics, as mentioned previously, there are two concerns. The first has to do with the specification of zero variation $\delta u = 0$ at the beginning and end of the time interval. As noted by Tonti (1973), by using this approach of Hamilton, one basically treats the initial value problem by constructing an artificial substitute boundary value problem. How can we justify fixing $u$ at the end of the time interval without variation, when this is the primary unknown of the initial value problem? Instead, due to knowledge of the initial conditions for displacement and velocity, should we not have $\delta u = 0$ and $\delta \dot{u} = 0$ at the beginning of the time interval? The second concern is the inability of Hamilton's principle to accommodate dissipation, which is the subject of the following two subsections.

*2.2. Rayleigh dissipation*

Since the original formulation by Hamilton (1834, 1835), there have been numerous attempts to extend the Lagrangian methodology to dynamical systems with dissipation. In perhaps the most useful of these efforts, Rayleigh (1877) introduced a separate function $F$ for dissipative elements.

As an illustration of this approach, let us consider the damped linear system displayed in Figure 2, governed by the equation of motion

$$m\ddot{u} + c\dot{u} + ku = \hat{f} \tag{14}$$

along with the initial conditions

$$u(0) = u_0; \quad \dot{u}(0) = v_0 \tag{15a,b}$$

For this system, we may define a dissipation function in the following form:

$$F(\dot{u};t) = \tfrac{1}{2} c \left[\dot{u}(t)\right]^2 \qquad (16)$$

in addition to the Lagrangian specified in (4) and (5). Although the action itself no longer can be written in explicit form, the first variation of $I$ is defined as follows:

$$\delta I = \delta \int_0^t L(u,\dot{u};\tau) d\tau - \int_0^t \frac{\partial F(\dot{u};\tau)}{\partial \dot{u}} \delta u \, d\tau \qquad (17)$$

or

$$\delta I = \int_0^t \left[ \frac{\partial L}{\partial \dot{u}} \delta \dot{u} + \frac{\partial L}{\partial u} \delta u - \frac{\partial F}{\partial \dot{u}} \delta u \right] d\tau = 0 \qquad (18)$$

Then, from (4), (5) and (16),

$$\delta I = \int_0^t \left[ m\dot{u}\delta\dot{u} - ku\delta u + \hat{f}\delta u - c\dot{u}\delta u \right] d\tau = 0 \qquad (19)$$

After applying integration-by-parts on the first term and assuming arbitrary variations of $\delta u$ over the interval, along with $\delta u(0) = \delta u(t) = 0$, we recover the equation of motion (14) for the damped oscillator as the Euler-Lagrange equation.

While this approach can lead to the proper governing differential equation of motion, it is not entirely satisfactory as a variational statement. In particular, one can no longer construct an action in explicit form and the dissipation function enters the first variation (17) in an *ad hoc* manner. Once again, the variation of displacement must be assumed zero at the end of the time interval, as in the original Hamilton's principle for the conservative system, and there is nothing that can be said about the initial conditions specified in (15).

*2.3. Other approaches to incorporate dissipation*

Many other formulations have been proposed in the ensuing years to define a Lagrangian for dissipative systems, although none are entirely satisfactory. To set the stage for the review of this work, perhaps we should begin with a corollary presented by Bauer (1931). Based upon the fundamental ideas in Davis (1928) and Morse (1930) on inverse problems in the calculus of variations and self-adjoint systems, Bauer developed two theorems and a corollary that states "The equations of motion of a dissipative linear dynamical system with constant coefficients are not given by a variational principle." The truth of this statement, as we shall see, depends upon the restrictions placed upon the definition of a variational principle for a given system. For example, Bateman (1931) introduced additional variables that produce a set of complementary Euler-Lagrange equations, which include the desired dissipative equation of motion. For instance, by selecting a Lagrangian in the form:

$$L(u,\dot{u},\ddot{u},v;t) = m\ddot{u}v + c\dot{u}v + kuv - \hat{f}(u+v) \tag{20}$$

with $v$ as an auxiliary variable, one can indeed obtain the equation of motion (14) as an Euler-Lagrange equation. However, now the Lagrangian involves the second time derivative of $u$ and the variational process also provides the extraneous relation

$$m\ddot{v} - c\dot{v} + kv = \hat{f} \tag{21}$$

along with supplemental conditions requiring the variations $\delta u$ and $\delta \dot{u}$ both to be zero at the beginning and end of the time interval. Clearly, (21) provides the governing equation for a related fictitious system, having negative damping. Several decades later, Morse and Feshbach (1953) adopt this device with some reservation in their classical text to accommodate dissipation within a variational framework. For the forced damped oscillator of Figure 2, their Lagrangian would be written in the somewhat more symmetrical form:

$$L(u,\dot{u},v,\dot{v};t) = m\dot{u}\dot{v} - \tfrac{1}{2}c(\dot{u}v - u\dot{v}) - kuv + \hat{f}(u+v) \tag{22}$$

This again leads to (14) and (21) as its Euler-Lagrange equations, along with supplemental conditions requiring zero variations of $\delta u$ and $\delta v$ at the beginning and end of the time interval. We should point out that neither Bateman (1931) nor Morse and Feshbach (1953) give any mention of the required supplemental conditions on the variations. By this time, the fixed endpoint concept of Hamilton had been fully ingrained in their thinking.

An alternative to the above approach is to assume that the dissipation comes from the fine scale interaction of infinitely many oscillators and then to use statistical methods to capture the dissipation through ensemble averaging (*e.g.*, Lanczos, 1949). However, this seems a long way to go to handle the simple damped oscillator. A more interesting and direct approach was introduced by Riewe (1996, 1997). He argues that the generalization of mechanics to include fractional derivatives enables the representation of dissipation within a variational framework and thus the limitations specified in the Bauer corollary are circumvented by permitting fractional order derivatives to appear in the Lagrangian. In particular, Riewe proposes to represent the dissipative effects with semi-derivatives (*i.e.*, derivatives of one-half order). While the idea has merit, the dissipative terms presented by Riewe (1996, 1997) do not solve the problem and, in fact, reduce to zero in most cases, as will be demonstrated below in Section 4 through a general result for fractional integration-by-parts. More recently, Cresson (2007) and others have attempted to correct the deficiencies, for example, by expanding the function space. This cannot be helpful, however, because we know very well that the exact solutions for many damped oscillator initial value problems are analytic. Again, we find in all of this work that attention is focused exclusively on reproducing the governing differential equations, with no concern for the supplemental endpoint constraints or for including the initial conditions within the variational formulation.

## 3. Convolution-based approaches for single degree-of-freedom oscillators

### 3.1. Convolution operators

In this section, we examine work that explicitly addresses the endpoint inconsistencies noted above and brings us closer to a true variational formulation for dynamics, including a representation for dissipative effects. These approaches developed by Gurtin (1964a,b) and Tonti (1973) are based on convolutions, where we define the convolution $[u*v](t)$ of two Lebesgue integrable functions $u(t)$ and $v(t)$ over the time interval from $0$ to $t$ by the relation:

$$[u*v](t) = \int_0^t u(\tau)v(t-\tau)d\tau \tag{22}$$

Then, convolution is seen to represent a bilinear form with the properties of commutativity

$$[u*v](t) = [v*u](t) \tag{23}$$

associativity

$$[u*[v*w]](t) = [[u*v]*w](t) \tag{24}$$

and distributivity

$$[u*[v+w]](t) = [u*v](t) + [u*w](t) \tag{25}$$

where $w(t)$ also is assumed Lebesgue integrable.

*3.2 Gurtin formalism*

In a pair of papers, Gurtin (1964a,b) developed new variational principles for the prototypical partial differential equations associated with wave and heat phenomena and for elastodynamics, respectively, based upon the concept of convolution operators. He is perhaps the first to recognize the inconsistency of Hamilton's principle for initial value problems, although he restricts his attention and criticism to continuum systems only.

Let us now formulate a Gurtin variational principle for the discrete damped oscillator governed by (14) and (15), as an extension of the work developed in Gurtin (1964a). We begin by transforming the initial value problem into the Laplace transform domain to obtain the following:

$$s^2 m\tilde{u} - smu_0 - mv_0 + sc\tilde{u} - cu_0 + k\tilde{u} = \tilde{f} \tag{26}$$

Here $\tilde{u} = \tilde{u}(s)$ and $\tilde{f} = \tilde{f}(s)$ are the Laplace transforms of $u(t)$ and $\hat{f}(t)$, respectively, with $s$ as the transform parameter. Then, dividing through by $s^2$, we have

$$m\tilde{u} + \frac{c}{s}\tilde{u} + \frac{k}{s^2}\tilde{u} = \frac{1}{s^2}\tilde{f} + \frac{m}{s}u_0 + \frac{c}{s^2}u_0 + \frac{m}{s^2}v_0 \tag{27}$$

After performing an inverse Laplace transform, the initial value problem can be restated in the time domain in the form of the following integro-differential equation:

$$mu(t) + [c*u](t) + [kt*u](t) = f(t) \tag{28}$$

with

$$f(t) = [t*\hat{f}](t) + (m+ct)u_0 + mtv_0 \tag{29}$$

Based upon a proof given by Gurtin (1964a) for the wave problem, here $u(t)$ is a solution of the original initial value problem defined in (14) and (15a,b), if and only if it also is a solution to (28).

Having reformulated the initial value problem, based upon the ideas in Gurtin (1964a), we can next propose an action functional in the following form built upon the convolution operator:

$$I(u;t) = \tfrac{1}{2}m[u*u](t) + \tfrac{1}{2}[c*[u*u]](t) + \tfrac{1}{2}[kt*[u*u]](t) - [f*u](t) \tag{30}$$

Then, for stationary action, we must have

$$\delta I(u;t) = [\delta u * mu](t) + [\delta u * [c*u]](t) + [\delta u * [kt*u]](t) - [\delta u * f](t) = 0 \tag{31}$$

For arbitrary variations $\delta u(t)$ throughout the time interval $[0,t]$, we finally recover (28). Thus, this Gurtin-type variational principle encapsulates both the governing equation of motion (14) and the specified initial conditions (15a,b) by producing an equivalent integro-differential Euler-Lagrange equation. Moreover, as can be seen through this example, the approach can accommodate dissipative elements within a variational framework.

As mentioned previously, this approach was used by Gurtin to provide a variational statement for the wave and heat equations (Gurtin, 1964a) and for elastodynamics (Gurtin, 1964b). The approach is somewhat cumbersome because of the step involving the division by $s^2$ in the transform domain, which implies a double integration in the time domain and the conversion of the original differential equation into an integro-differential equation. Furthermore, it appears that the variation $\delta u$ must remain completely arbitrary, with no restriction on the variations of $u$ and $\dot{u}$ at time zero, despite their specification as part of the initial value problem definition. Thus, while a Gurtin formulation does incorporate the initial conditions into the variational framework, it ignores the requirement to limit the variations accordingly.

*3.3. Tonti formalism*

Tonti (1973) provides strong arguments based upon theoretical concepts from functional analysis that Hamilton's principle addresses a substitute boundary value problem, rather than the intended initial value problem. In particular, Tonti shows that the bilinear form used in Hamilton's principle to construct the action, such as in (6) above, is formally self-adjoint for a range of conservative dynamical systems, meaning that it produces the correct differential equations as its Euler-Lagrange equations. However, this bilinear form is not self-adjoint with respect to the initial conditions, requiring instead the specification of the function at the beginning and end of the time interval, as we have already discussed. In its place, Tonti advocates that a convolutional bilinear form should be adopted as the basis for the development of variational principles in dynamics. This is consistent with the method proposed by Gurtin, although Tonti (1973) presents a broader view, while proposing a more simplified approach.

Let us first consider a damped oscillator initial value problem defined by (14) and homogeneous initial conditions, using the Tonti formalism. By extending the applications examined in Tonti (1973), we may write an action with terms involving convolutions to represent the inertial, dissipative and restoring elements in the system. Thus, we let

$$I(u;t) = \tfrac{1}{2}\int_0^t \frac{du(t-\tau)}{d(t-\tau)} m \frac{du(\tau)}{d\tau} d\tau + \tfrac{1}{2}\int_0^t \frac{du(t-\tau)}{d(t-\tau)} cu(\tau)d\tau$$
$$+ \tfrac{1}{2}\int_0^t u(t-\tau)ku(\tau)d\tau - \int_0^t u(t-\tau)\hat{f}(\tau)d\tau \quad (32)$$

where we choose to write the integrands in explicit detail, using an order that later will permit us to extend the formulation to multi-degree-of-freedom dynamical systems in a straightforward manner. Then, for a stationary point of this action, we must have

$$\delta I(u;t) = \int_0^t \frac{d\delta u(t-\tau)}{d(t-\tau)} m \frac{du(\tau)}{d\tau} d\tau + \tfrac{1}{2}\int_0^t \frac{d\delta u(t-\tau)}{d(t-\tau)} cu(\tau)d\tau + \tfrac{1}{2}\int_0^t \delta u(t-\tau) c \frac{du(\tau)}{d\tau} d\tau$$
$$+ \int_0^t \delta u(t-\tau)ku(\tau)d\tau - \int_0^t \delta u(t-\tau)\hat{f}(\tau)d\tau = 0 \quad (33)$$

However, from integration-by-parts on the inertial contribution

$$\int_0^t \frac{d\delta u(t-\tau)}{d(t-\tau)} m \frac{du(\tau)}{d\tau} d\tau = \int_0^t \frac{d\delta u(\tau)}{d\tau} m \frac{du(t-\tau)}{d(t-\tau)} d\tau$$
$$= \int_0^t \frac{d}{d\tau}\left[\delta u(\tau) m \frac{du(t-\tau)}{d(t-\tau)}\right] d\tau - \int_0^t \delta u(\tau) m \frac{d}{d\tau}\frac{du(t-\tau)}{d(t-\tau)} d\tau$$
$$= \left[\delta u(\tau) m \frac{du(t-\tau)}{d(t-\tau)}\right]_0^t + \int_0^t \delta u(\tau) m \frac{d^2 u(t-\tau)}{d(t-\tau)^2} d\tau \quad (34)$$
$$= -\left[\delta u(t-\tau) m \frac{du(\tau)}{d\tau}\right]_0^t + \int_0^t \delta u(t-\tau) m \frac{d^2 u(\tau)}{d\tau^2} d\tau$$

Similarly, for the first damping term, after some manipulation, we find

$$\tfrac{1}{2}\int_0^t \frac{d\delta u(t-\tau)}{d(t-\tau)} cu(\tau)d\tau = -\tfrac{1}{2}\big[\delta u(t-\tau)cu(\tau)\big]_0^t + \tfrac{1}{2}\int_0^t \delta u(t-\tau) c \frac{du(\tau)}{d\tau} d\tau \quad (35)$$

Substituting (34) and (35) into (33) and then rearranging terms yields

$$\delta I(u;t) = \int_0^t \delta u(t-\tau) \left[ m\frac{d^2 u(\tau)}{d\tau^2} + c\frac{du(\tau)}{d\tau} + ku(\tau) - \hat{f}(\tau) \right] d\tau$$
$$- \left[ \delta u(t-\tau) \left( m\frac{du(\tau)}{d\tau} + \tfrac{1}{2} cu(\tau) \right) \right]_0^t = 0 \quad (36)$$

Now, we consider arbitrary variations, except for the constraint $\delta u(0) = 0$. As a result, (36) provides the following Euler-Lagrange equations:

$$m\frac{d^2 u(\tau)}{d\tau^2} + c\frac{du(\tau)}{d\tau} + ku(\tau) = \hat{f}(\tau) \quad \text{for } 0 \leq \tau \leq t \quad (37a)$$

$$m\frac{du(\tau)}{d\tau} + \tfrac{1}{2} cu(\tau) = 0 \quad \text{for } \tau = 0 \quad (37b)$$

representing the damped equation of motion and a non-physical initial condition on velocity and displacement, respectively. Notice also there is no constraint on the variation $\delta \dot{u}(0)$ that one would expect due to the specification of initial velocity.

Finally, we should mention that Arthurs and Jones (1976) attempted to correct the deficiencies in the Tonti (1973) formulation by introducing a number of additional initial condition terms. However, their formulation requires the inclusion of dissipation and ultimately involves completely free variations over the interval $[0,t]$, which is inconsistent with the known initial values of displacement and velocity.

To summarize this section, we should say that both Gurtin (1964a,b) and Tonti (1973) made important strides towards the establishment of a proper variational statement for dynamical systems. However, inconsistencies remain in both of these approaches, which suggest that further consideration is needed. In the following section, we provide some basic concepts related to fractional calculus that will prove relevant in this regard.

## 4. Fractional calculus concepts and integral relations

### 4.1. Basic definitions

The idea of generalizing derivatives to fractional order originated with l'Hôpital**,** Leibniz and their contemporaries shortly after the introduction of the calculus. A thorough historical review including this earliest work has been developed by Ross (1977), while the monographs by Oldham and Spanier (1974) and especially Samko *et al.* (1998) provide rather comprehensive collections of the known results. From this work, a number of different formulations for fractional integrals and derivatives emerge. However, here we will focus on the version attributed to Riemann and Liouville.

Following Samko *et al.* (1993), let $u(\tau)$ represent an $L_1$ Lebesgue integrable function over the interval $(0,t)$. Thus, $u(\tau) \in L_1(0,t)$ and for a fractional integral of order $\alpha$, we may write

$$\left(\mathcal{I}_{0^+}^\alpha u\right)(\tau) \equiv \frac{1}{\Gamma(\alpha)} \int_0^\tau \frac{u(\xi)}{(\tau-\xi)^{1-\alpha}} d\xi \quad \text{for } \tau > 0, \ \alpha > 0 \tag{38a}$$

$$\left(\mathcal{I}_{t^-}^\alpha u\right)(\tau) \equiv \frac{1}{\Gamma(\alpha)} \int_\tau^t \frac{u(\xi)}{(\xi-\tau)^{1-\alpha}} d\xi \quad \text{for } \tau < t, \ \alpha > 0 \tag{38b}$$

Here (38a) and (38b) represent the left and right Riemann-Liouville fractional integral of order $\alpha$, respectively, where we have chosen the left fractional integral to operate over the interval from 0 to $\tau$, while the right fractional integral works over the range from $\tau$ to $t$. From these definitons, we may recognize that fractional integration also can be viewed as a convolution of an $L_1$ function, say $u(\tau)$, with the kernel $\tau^{\alpha-1}/\Gamma(\alpha)$.

The fractional integration of such functions satisfies the following composition properties:

$$\left(\mathcal{I}_{0^+}^\alpha \left(\mathcal{I}_{0^+}^\beta u\right)\right)(\tau) = \left(\mathcal{I}_{0^+}^{\alpha+\beta} u\right)(\tau) \tag{39a}$$

$$\left(\mathcal{I}_{t^-}^\alpha \left(\mathcal{I}_{t^-}^\beta u\right)\right)(\tau) = \left(\mathcal{I}_{t^-}^{\alpha+\beta} u\right)(\tau) \tag{39b}$$

for $0 < \tau < t$ with $\alpha > 0$, $\beta > 0$.

Next, we consider the corresponding left and right Riemann-Liouville fractional derivatives of order $\alpha$, which can be written, respectively, in the following form

$$\left(\mathcal{D}_{0^+}^\alpha u\right)(\tau) \equiv +\frac{1}{\Gamma(1-\alpha)} \frac{d}{d\tau} \int_0^\tau \frac{u(\xi)}{(\tau-\xi)^\alpha} d\xi \quad \text{for } 0 \leq \tau \leq t, \ 0 < \alpha < 1 \tag{40a}$$

$$\left(\mathcal{D}_{t^-}^\alpha u\right)(\tau) \equiv -\frac{1}{\Gamma(1-\alpha)} \frac{d}{d\tau} \int_\tau^t \frac{u(\xi)}{(\xi-\tau)^\alpha} d\xi \quad \text{for } 0 \leq \tau \leq t,\ 0 < \alpha < 1 \tag{40b}$$

These exist almost everywhere in the interval for absolutely continuous $u(\tau)$.

We also may develop an alternative expression for the left Riemann-Liouville fractional derivative by using the commutativity property (23) of the convolution operator in (40a), along with the Leibniz rule. With this approach, we find (Samko *et al*., 1993)

$$\begin{aligned}
\left(\mathcal{D}_{0^+}^\alpha u\right)(\tau) &= +\frac{1}{\Gamma(1-\alpha)} \frac{d}{d\tau} \int_0^\tau \frac{u(\xi)}{(\tau-\xi)^\alpha} d\xi \\
&= \frac{1}{\Gamma(1-\alpha)} \frac{d}{d\tau} \int_0^\tau \frac{u(\tau-\xi)}{(\xi)^\alpha} d\xi \\
&= \frac{1}{\Gamma(1-\alpha)} \frac{u(0)}{(\tau)^\alpha} + \frac{1}{\Gamma(1-\alpha)} \int_0^\tau \frac{\dot{u}(\tau-\xi)}{(\xi)^\alpha} d\xi \\
&= \frac{1}{\Gamma(1-\alpha)} \frac{u(0)}{(\tau)^\alpha} + \frac{1}{\Gamma(1-\alpha)} \int_0^\tau \frac{\dot{u}(\xi)}{(\tau-\xi)^\alpha} d\xi \\
&= \frac{1}{\Gamma(1-\alpha)} \frac{u(0)}{(\tau)^\alpha} + \left(\mathcal{J}_{0^+}^{1-\alpha} \dot{u}\right)(\tau)
\end{aligned} \tag{41a}$$

Similarly, by beginning with (40b), we can formulate an alternate expression for the right Riemann-Liouville fractional derivative through the following steps

$$\begin{aligned}
\left(\mathcal{D}_{t^-}^\alpha u\right)(\tau) &= -\frac{1}{\Gamma(1-\alpha)} \frac{d}{d\tau} \int_\tau^t \frac{u(\xi)}{(\xi-\tau)^\alpha} d\xi \\
&= -\frac{1}{\Gamma(1-\alpha)} \frac{d}{d\tau} \int_\tau^t \frac{u(\tau-\xi+t)}{(t-\xi)^\alpha} d\xi \\
&= \frac{1}{\Gamma(1-\alpha)} \frac{u(t)}{(t-\tau)^\alpha} - \frac{1}{\Gamma(1-\alpha)} \int_\tau^t \frac{\dot{u}(\tau-\xi+t)}{(t-\xi)^\alpha} d\xi \\
&= \frac{1}{\Gamma(1-\alpha)} \frac{u(t)}{(t-\tau)^\alpha} - \frac{1}{\Gamma(1-\alpha)} \int_\tau^t \frac{\dot{u}(\xi)}{(t-\tau)^\alpha} d\xi \\
&= \frac{1}{\Gamma(1-\alpha)} \frac{u(t)}{(t-\tau)^\alpha} - \left(\mathcal{J}_{t^-}^{1-\alpha} \dot{u}\right)(\tau)
\end{aligned} \tag{41b}$$

Combining the above definitions for any summable function $u(\tau)$, we find, in general, for the composition of left operators

$$\left(\mathcal{D}_{0^+}^\alpha\left(\mathcal{I}_{0^+}^\alpha u\right)\right)(\tau) = u(\tau) \tag{42a}$$

while by reversing the order, we have instead (Samko *et al.*, 1993)

$$\left(\mathcal{I}_{0^+}^\alpha\left(\mathcal{D}_{0^+}^\alpha u\right)\right)(\tau) = u(\tau) - \frac{1}{\Gamma(\alpha)}\frac{\left(\mathcal{I}_{0^+}^{1-\alpha}u\right)(0)}{\tau^{1-\alpha}} \tag{42b}$$

In a similar way, for the right operators, we may write

$$\left(\mathcal{D}_{t^-}^\alpha\left(\mathcal{I}_{t^-}^\alpha u\right)\right)(\tau) = u(\tau) \tag{43a}$$

and

$$\left(\mathcal{I}_{t^-}^\alpha\left(\mathcal{D}_{t^-}^\alpha u\right)\right)(\tau) = u(\tau) - \frac{1}{\Gamma(\alpha)}\frac{\left(\mathcal{I}_{t^-}^{1-\alpha}u\right)(t)}{(t-\tau)^{1-\alpha}} \tag{43b}$$

Notice that all four of these relations (42a), (42b), (43a) and (43b) approach the usual results from ordinary calculus, as we let $\alpha$ tend to unity from below.

*4.2. Integration-by-parts*

Integration-by-parts plays a central role in the development of variational principles of mechanics. In this section, we begin with the classical result by Love and Young (1938) for fractional integration-by-parts and then derive several other versions that will be useful in our subsequent development.

For functions $f(\tau) \in L_p$ and $g(\tau) \in L_q$ with $p \geq 1$, $q \geq 1$ and $1/p + 1/q \leq 1 + \alpha$, Love and Young (1938) develop a theorem for fractional integration-by-parts, based upon the work of Hardy and Littlewood (1928). Alternatively, the proof can be constructed from the general Fubini Theorem for the interchangeability of iterated integrals. Thus, with $f$ and $g$ defined, such that $\iint f(\xi)g(\zeta)|\xi-\zeta|^{\alpha-1}d\xi d\zeta$ is absolutely convergent over $\xi \geq 0$, $\zeta \geq 0$ and therefore over $0 \leq \xi \leq \zeta \leq t$, we can write for the inner product of $f$ with a left fractional integral of $g$

$$\int_0^t f(\tau)\left(\mathcal{I}_{0^+}^\alpha g\right)(\tau)d\tau = \frac{1}{\Gamma[\alpha]}\int_0^t f(\tau)\int_0^\tau g(\zeta)(\tau-\zeta)^{\alpha-1}d\zeta d\tau$$

$$= \frac{1}{\Gamma[\alpha]}\int_0^t\int_0^\tau f(\tau)g(\zeta)(\tau-\zeta)^{\alpha-1}d\zeta d\tau$$

$$= \frac{1}{\Gamma[\alpha]}\int_0^t\int_\zeta^t f(\tau)g(\zeta)(\tau-\zeta)^{\alpha-1}d\tau d\zeta \qquad (44)$$

$$= \frac{1}{\Gamma[\alpha]}\int_0^t g(\zeta)\int_\zeta^t f(\tau)(\tau-\zeta)^{\alpha-1}d\tau d\zeta$$

which, after an interchange of dummy variables, provides the theorem as

$$\int_0^t f(\tau)\left(\mathcal{I}_{0^+}^\alpha g\right)(\tau)d\tau = \int_0^t \left(\mathcal{I}_{t^-}^\alpha f\right)(\tau)g(\tau)d\tau \qquad (45)$$

Here we see that via the Love-Young integration-by-parts theorem, the left Riemann-Liouville fractional integral of one of a pair of functions within an inner product can be transferred to the other as a right Riemann-Liouville fractional integral of the same order.

Now we will begin to develop some new ideas that will be directly applicable to our formulation of variational principles of dynamics. We begin by considering (45) and letting

$$f(\tau) = \left(\mathcal{D}_{t^-}^\alpha \varphi\right)(\tau) \qquad (46a)$$

$$g(\tau) = \left(\mathcal{D}_{0^+}^\alpha \psi\right)(\tau) \qquad (46b)$$

Then, after substituting (46a) and (46b) into (45), we have

$$\int_0^t \left(\mathcal{D}_{t^-}^\alpha \varphi\right)(\tau)\left(\mathcal{I}_{0^+}^\alpha \left(\mathcal{D}_{0^+}^\alpha \psi\right)\right)(\tau)d\tau = \int_0^t \left(\mathcal{I}_{t^-}^\alpha \left(\mathcal{D}_{t^-}^\alpha \varphi\right)\right)(\tau)\left(\mathcal{D}_{0^+}^\alpha \psi\right)(\tau)d\tau \qquad (47)$$

By invoking (42b) and (43b), this can be rewritten

$$\int_0^t \left(\mathcal{D}_{t^-}^\alpha \varphi\right)(\tau)\left\{\psi(\tau) - \frac{1}{\Gamma(\alpha)}\frac{\left(\mathcal{I}_{0^+}^{1-\alpha}\psi\right)(0)}{\tau^{1-\alpha}}\right\}d\tau = \int_0^t \left\{\varphi(\tau) - \frac{1}{\Gamma(\alpha)}\frac{\left(\mathcal{I}_{t^-}^{1-\alpha}\varphi\right)(t)}{(t-\tau)^{1-\alpha}}\right\}\left(\mathcal{D}_{0^+}^\alpha \psi\right)(\tau)d\tau \qquad (48)$$

However,

$$\int_0^t \left(\mathcal{D}_{t^-}^\alpha \varphi\right)(\tau) \frac{1}{\Gamma(\alpha)} \frac{\left(\mathcal{I}_{0^+}^{1-\alpha}\psi\right)(0)}{\tau^{1-\alpha}} d\tau = \left(\mathcal{I}_{0^+}^{1-\alpha}\psi\right)(0) \frac{1}{\Gamma(\alpha)} \int_0^t \frac{\left(\mathcal{D}_{t^-}^\alpha \varphi\right)(\tau)}{\tau^{1-\alpha}} d\tau$$

$$= \left(\mathcal{I}_{0^+}^{1-\alpha}\psi\right)(0)\left(\mathcal{I}_{t^-}^\alpha\left(\mathcal{D}_{t^-}^\alpha \varphi\right)\right)(0) \quad (49a)$$

$$= \left(\mathcal{I}_{0^+}^{1-\alpha}\psi\right)(0)\left\{\varphi(0) - \frac{1}{\Gamma(\alpha)} \frac{\left(\mathcal{I}_{t^-}^{1-\alpha}\varphi\right)(t)}{(t)^{1-\alpha}}\right\}$$

and

$$\int_0^t \frac{1}{\Gamma(\alpha)} \frac{\left(\mathcal{I}_{t^-}^{1-\alpha}\varphi\right)(t)}{(t-\tau)^{1-\alpha}} \left(\mathcal{D}_{0^+}^\alpha \psi\right)(\tau) d\tau = \left(\mathcal{I}_{t^-}^{1-\alpha}\varphi\right)(t) \frac{1}{\Gamma(\alpha)} \int_0^t \frac{\left(\mathcal{D}_{0^+}^\alpha \psi\right)(\tau)}{(t-\tau)^{1-\alpha}} d\tau$$

$$= \left(\mathcal{I}_{t^-}^{1-\alpha}\varphi\right)(t)\left(\mathcal{I}_{0^+}^\alpha\left(\mathcal{D}_{0^+}^\alpha \psi\right)\right)(\tau) \quad (49b)$$

$$= \left(\mathcal{I}_{t^-}^{1-\alpha}\varphi\right)(t)\left\{\psi(t) - \frac{1}{\Gamma(\alpha)} \frac{\left(\mathcal{I}_{0^+}^{1-\alpha}\psi\right)(0)}{t^{1-\alpha}}\right\}$$

After substituting (49a) and (49b) into (48), cancelling the common term and rearranging the result, we arrive at the following general integration-by-parts relation for fractional derivatives:

$$\int_0^t \varphi(\tau)\left(\mathcal{D}_{0^+}^\alpha \psi\right)(\tau) d\tau = \int_0^t \left(\mathcal{D}_{t^-}^\alpha \varphi\right)(\tau)\psi(\tau) d\tau + \left(\mathcal{I}_{t^-}^{1-\alpha}\varphi\right)(t)\psi(t) - \varphi(0)\left(\mathcal{I}_{0^+}^{1-\alpha}\psi\right)(0) \quad (50)$$

Here, we begin with the inner product of one function with the left fractional derivative of another and then end with the inner product of the latter function with a right fractional derivative of the former function, plus some boundary terms involving both left and right fractional integrals.

We can easily see that the usual integration-by-parts result is obtained by passing to the limit $\alpha \to 1$, in which case (50) approaches the well-known relation

$$\int_0^t \varphi(\tau)\frac{d\psi}{d\tau}(\tau) d\tau = -\int_0^t \frac{d\varphi}{d\tau}(\tau)\psi(\tau) d\tau + \varphi(t)\psi(t) - \varphi(0)\psi(0) \quad (51)$$

while for the other limit as $\alpha \to 0$, (50) reduces to a simple tautology.

*4.3. Reflections and convolutions*

We now extend the above relations for fractional calculus by considering reflections and convolutions. Let $\hat{g}$ represent the reflection of $g$, such that $\hat{g}(\tau) = g(t-\tau)$. Then,

$$\begin{aligned}\left(\mathcal{I}_{t^-}^\alpha \hat{g}\right)(\tau) &= \frac{1}{\Gamma(\alpha)} \int_\tau^t \frac{\hat{g}(\xi)}{(\xi-\tau)^{1-\alpha}} d\xi \\ &= \frac{1}{\Gamma(\alpha)} \int_\tau^t \frac{g(t-\xi)}{(\xi-\tau)^{1-\alpha}} d\xi \\ &= \frac{1}{\Gamma(\alpha)} \int_0^{t-\tau} \frac{g(\gamma)}{((t-\tau)-\gamma)^{1-\alpha}} d\gamma \\ &= \left(\mathcal{I}_{0^+}^\alpha g\right)(t-\tau)\end{aligned} \qquad (52a)$$

Similarly, we find

$$\left(\mathcal{I}_{0^+}^\alpha \hat{g}\right)(\tau) = \left(\mathcal{I}_{t^-}^\alpha g\right)(t-\tau) \qquad (52b)$$

From (52a), a left fractional convolutional integration-by-parts can be developed as follows:

$$\begin{aligned}\int_0^t f(\tau)\left(\mathcal{I}_{0^+}^\alpha g\right)(t-\tau)d\tau &= \int_0^t f(\tau)\left(\mathcal{I}_{t^-}^\alpha \hat{g}\right)(\tau)d\tau \\ &= \int_0^t \left(\mathcal{I}_{0^+}^\alpha f\right)(\tau)\hat{g}(\tau)d\tau \\ &= \int_0^t \left(\mathcal{I}_{0^+}^\alpha f\right)(\tau)g(t-\tau)d\tau\end{aligned} \qquad (53a)$$

while for right fractional convolutional integration-by-parts, by using (52b), we have

$$\int_0^t f(\tau)\left(\mathcal{I}_{t^-}^\alpha g\right)(t-\tau)d\tau = \int_0^t \left(\mathcal{I}_{t^-}^\alpha f\right)(\tau)g(t-\tau)d\tau \qquad (53b)$$

Notice that for convolutions, the Riemann-Liouville fractional integral of one of a pair of functions is transferred to the other as a fractional integral of the same side (*i.e.*, left or right) and order. Recall from (45) that for integration-by-parts with inner products, the sidedness reverses from left to right or *vice versa*.

Next, we assume

$$f(\tau) = \left(\mathcal{D}_{0^+}^\alpha \varphi\right)(\tau) \qquad (54a)$$

$$g(\tau) = \left(\mathcal{D}_{0^+}^{\alpha}\psi\right)(\tau) \tag{54b}$$

and substitute these expressions into (53a) to yield

$$\int_0^t \left(\mathcal{D}_{0^+}^{\alpha}\varphi\right)(\tau)\left(\mathcal{I}_{0^+}^{\alpha}\left(\mathcal{D}_{0^+}^{\alpha}\psi\right)\right)(t-\tau)d\tau = \int_0^t \left(\mathcal{I}_{0^+}^{\alpha}\left(\mathcal{D}_{0^+}^{\alpha}\varphi\right)\right)(\tau)\left(\mathcal{D}_{0^+}^{\alpha}\psi\right)(t-\tau)d\tau \tag{55}$$

Then, after invoking (42b) on both sides of (55) and performing a few straightforward manipulations, we obtain the following general integration-by-parts relation for convolution with left fractional derivatives:

$$\int_0^t \varphi(\tau)\left(\mathcal{D}_{0^+}^{\alpha}\psi\right)(t-\tau)d\tau = \int_0^t \left(\mathcal{D}_{0^+}^{\alpha}\varphi\right)(\tau)\psi(t-\tau)d\tau + \left(\mathcal{I}_{0^+}^{1-\alpha}\varphi\right)(0)\psi(t) - \varphi(t)\left(\mathcal{I}_{0^+}^{1-\alpha}\psi\right)(0) \tag{56}$$

In a similar manner, we may substitute

$$f(\tau) = \left(\mathcal{D}_{t^-}^{\alpha}\varphi\right)(\tau) \tag{57a}$$

$$g(\tau) = \left(\mathcal{D}_{t^-}^{\alpha}\psi\right)(\tau) \tag{57b}$$

into (53b) to produce

$$\int_0^t \left(\mathcal{D}_{t^-}^{\alpha}\varphi\right)(\tau)\left(\mathcal{I}_{t^-}^{\alpha}\left(\mathcal{D}_{t^-}^{\alpha}\psi\right)\right)(t-\tau)d\tau = \int_0^t \left(\mathcal{I}_{t^-}^{\alpha}\left(\mathcal{D}_{t^-}^{\alpha}\varphi\right)\right)(\tau)\left(\mathcal{D}_{t^-}^{\alpha}\psi\right)(t-\tau)d\tau \tag{58}$$

Then, invoking (43b) on both sides and carrying out a few now familiar steps, we may write

$$\int_0^t \varphi(\tau)\left(\mathcal{D}_{t^-}^{\alpha}\psi\right)(t-\tau)d\tau = \int_0^t \left(\mathcal{D}_{t^-}^{\alpha}\varphi\right)(\tau)\psi(t-\tau)d\tau + \left(\mathcal{I}_{t^-}^{1-\alpha}\varphi\right)(t)\psi(0) - \varphi(0)\left(\mathcal{I}_{t^-}^{1-\alpha}\psi\right)(t) \tag{59}$$

as the corresponding integration-by-parts formula involving convolution with right fractional derivatives.

Two points should be made. First, we see that, for convolutional integration-by-parts, the sidedness (*i.e.*, left or right) of all of the fractional derivatives and integrals remain the same, unlike the case for inner products. Second, the ordinary integration-by-parts formula for convolution

$$\int_0^t \varphi(t-\tau)\frac{d\psi}{d\tau}(\tau)d\tau = \int_0^t \frac{d\varphi}{d\tau}(\tau)\psi(t-\tau)d\tau + \varphi(0)\psi(t) - \varphi(t)\psi(0) \tag{60}$$

is recovered by approaching the limit $\alpha \to 1$ for both (56) and (59).

*4.4. Complementary order fractional derivatives*

In (50), (56) and (59), through the process of integration-by-parts, we may notice the appearance of fractional integrals of order $1-\alpha$, which we will denote as complementary to order $\alpha$. Let us now examine several special cases, involving the inner product and convolution of complementary order fractional derivatives.

As a first case, we will consider the inner product integration-by-parts relation by substituting

$$\varphi(\tau) = \left(\mathcal{D}_{t^-}^{1-\alpha} u\right)(\tau) \tag{61a}$$

$$\psi(\tau) = u(\tau) \tag{61b}$$

into (50) for a summable function $u(\tau)$ over the interval $(0,t)$. Then, for $0 < \alpha < 1$, using integration-by-parts on the left fractional derivative that appears in the left-hand side integral, we have

$$\int_0^t \left(\mathcal{D}_{t^-}^{1-\alpha} u\right)(\tau)\left(\mathcal{D}_{0^+}^{\alpha} u\right)(\tau)d\tau = \int_0^t \left(\mathcal{D}_{t^-}^{\alpha} \mathcal{D}_{t^-}^{1-\alpha} u\right)(\tau) u(\tau) d\tau$$
$$+ \left(\mathcal{I}_{t^-}^{1-\alpha} \mathcal{D}_{t^-}^{1-\alpha} u\right)(t) u(t) - \left(\mathcal{D}_{t^-}^{1-\alpha} u\right)(0)\left(\mathcal{I}_{0^+}^{1-\alpha} u\right)(0) \tag{62a}$$
$$= \int_0^t \left(\mathcal{D}_{t^-}^{1} u\right)(\tau) u(\tau) d\tau + u(t)u(t)$$

Similarly, by shifting the right fractional derivative in the left-hand side integral, we may write

$$\int_0^t \left(\mathcal{D}_{t^-}^{1-\alpha} u\right)(\tau)\left(\mathcal{D}_{0^+}^{\alpha} u\right)(\tau)d\tau = \int_0^t u(\tau)\left(\mathcal{D}_{0^+}^{1-\alpha} \mathcal{D}_{0^+}^{\alpha} u\right)(\tau) d\tau$$
$$- \left(\mathcal{I}_{t^-}^{\alpha} u\right)(t)\left(\mathcal{D}_{0^+}^{\alpha} u\right)(t) + u(0)\left(\mathcal{I}_{0^+}^{\alpha} \mathcal{D}_{0^+}^{\alpha} u\right)(0) \tag{62b}$$
$$= \int_0^t u(\tau)\left(\mathcal{D}_{0^+}^{1} u\right)(\tau) d\tau + u(0)u(0)$$

Then, after recognizing that $\left(\mathcal{D}_{t^-}^{1} u\right)(\tau) = -\dot{u}(\tau)$ and $\left(\mathcal{D}_{0^+}^{1} u\right)(\tau) = \dot{u}(\tau)$ and equating the right-hand sides of (62a) and (62b), we find

$$-\int_0^t \dot{u}(\tau) u(\tau) d\tau + u^2(t) = \int_0^t u(\tau) \dot{u}(\tau) d\tau + u^2(0) \tag{63}$$

This leads to the well-known classical result

$$\int_0^t u(\tau) \dot{u}(\tau) d\tau = \tfrac{1}{2}\left[u^2(t) - u^2(0)\right] \tag{64}$$

which indicates that the inner product of displacement and velocity is path independent and therefore not a useful representation for damping. Of course, this is why one cannot write a dissipative Lagrangian using ordinary calculus.

Now, if we substitute (64) into the right-hand side of either (62a) or (62b), we find that for $0 < \alpha < 1$ the inner product of complementary fractional derivatives of displacement reduces to the following form:

$$\int_0^t \left(\mathcal{D}_{t^-}^{1-\alpha} u\right)(\tau)\left(\mathcal{D}_{0^+}^{\alpha} u\right)(\tau) d\tau = \tfrac{1}{2}\left[u^2(t) + u^2(0)\right] \tag{65}$$

where again the integral is path independent. As a result, this generalized form of the Lagrangian term proposed by Riewe (1995, 1996), Cresson (2007) and others also cannot be used to model damping.

Additionally, we should note the difference in the right-hand sides between (64) and (65), which are caused by the subtle memoryless character of integer order derivatives. This can be seen clearly by considering the case with constant $u(\tau) = u_0$ throughout the time interval. Then, by including the first derivative of the function in the integrand in (64), we lose all information about the signal and return zero for the integral. On the other hand, for $0 < \alpha < 1$, the fractional derivatives appearing in the integrand in (65) retain a sense of the entire history of the signal $u(\tau)$ and return for this integral the value $u_0^2$.

Let us next examine the convolution of complementary fractional derivatives. For this we take

$$\varphi(\tau) = \left(\mathcal{D}_{0^+}^{1-\alpha} u\right)(\tau) \tag{66a}$$

$$\psi(\tau) = u(\tau) \tag{66b}$$

and substitute into (56) to provide

$$\int_0^t \left(\mathcal{D}_{0^+}^{1-\alpha} u\right)(\tau)\left(\mathcal{D}_{0^+}^{\alpha} u\right)(t-\tau) d\tau = \int_0^t \left(\mathcal{D}_{0^+}^{\alpha} \mathcal{D}_{0^+}^{1-\alpha} u\right)(\tau) u(t-\tau) d\tau$$

$$+ \left(\mathcal{J}_{0^+}^{1-\alpha} \mathcal{D}_{0^+}^{1-\alpha} u\right)(0) u(t) - \left(\mathcal{D}_{0^+}^{1-\alpha} u\right)(t)\left(\mathcal{J}_{0^+}^{1-\alpha} u\right)(0) \tag{67}$$

$$= \int_0^t \dot{u}(\tau) u(t-\tau) d\tau + u(0) u(t)$$

$0 < \alpha < 1$. An identical result is obtained, if we instead shift the complementary derivative via an integration-by-parts operation. Thus, we find that the convolution of complementary fractional derivatives retains path dependence. This suggests its potential applicability for describing dissipative processes, as will be discussed below in Section 5.

As a final result, we generalize (67) to operate on the fractional derivatives of two distinct summable functions $\varphi(\tau)$ and $\psi(\tau)$. For this case, we may write the integration-by-parts formula as

$$\int_0^t \left(\mathcal{D}_{0^+}^{1-\alpha}\varphi\right)(\tau)\left(\mathcal{D}_{0^+}^{\alpha}\psi\right)(t-\tau)d\tau = \int_0^t \frac{d\varphi}{d\tau}(\tau)\psi(t-\tau)d\tau + \varphi(0)\psi(t) \tag{68a}$$

The corresponding convolved integration-by-parts relationship for integral derivatives, which has been given previously in (60), is repeated here for comparative purposes

$$\int_0^t \varphi(t-\tau)\frac{d\psi}{d\tau}(\tau)d\tau = \int_0^t \frac{d\varphi}{d\tau}(\tau)\psi(t-\tau)d\tau + \varphi(0)\psi(t) - \varphi(t)\psi(0) \tag{68b}$$

Notice the difference in the released boundary terms between (68a) and (68b). Both forms will be used when we consider the stationarity of convolved action functionals in the following sections.

## 5. Principle of mixed convolved action for single degree of freedom oscillators

Recent work by Sivaselvan and Reinhorn (2006) and Sivaselvan *et al.* (2009) indicates a number of computational and theoretical benefits resulting from the adoption of a mixed Lagrangian formulation for structural dynamics. Here we continue along that line, but propose instead a *mixed convolved action*. In the current section, we focus on the single degree of freedom damped oscillator, displayed in Figure 2.

For the mixed formulation, let us introduce the displacement $u(t)$ and the impulse of the spring force $J(t)$ as the primary variables. The velocity $v(t)$ is then written as in (1), while the spring force becomes

$$F(t) = \dot{J}(t) \tag{69}$$

In terms of these mixed variables, the governing differential equations for the damped oscillator can be written

$$m\ddot{u} + c\dot{u} + \dot{J} = \hat{f} \tag{70a}$$

$$a\ddot{J} - \dot{u} = 0 \tag{70b}$$

where (70a) and (70b) represent the equations of motion and compatibility, respectively, valid at any time $t$. In addition to (70), the strong form of our problem includes the following initial conditions in mixed variables

$$m\dot{u}(0) + cu(0) + J(0) = \hat{j}(0) \tag{71a}$$

$$a\dot{J}(0) - u(0) = 0 \tag{71b}$$

where $\hat{j}(0)$ is the impulse of applied force $\hat{f}$ evaluated at time zero. Notice that in both (70b) and (71b), we work directly with the spring flexibility $a$, rather than the stiffness, where $a = 1/k$.

Our objective is to write an action that recovers all of these governing relations as the Euler-Lagrange equations for the damped oscillator. This action will include terms involving both ordinary first-order derivatives and fractional half-order or semi-derivatives. For convenience, in the following, we use the simplified notation $\breve{\varphi}$ to indicate the left Riemann-Liouville semi-derivative of $\varphi$. Thus, for example, we let

$$\breve{\varphi}(t) = \left(\mathcal{D}_{0^+}^{1/2}\varphi\right)(t) \tag{72}$$

Now with (70) and (71) as the representation of the initial value problem in strong form, let us propose the following *mixed convolved action* functional:

$$I_C(u,\breve{u},\dot{u},J,\breve{J},\dot{J};t) = \tfrac{1}{2}m\dot{u}*\dot{u} - \tfrac{1}{2}a\dot{J}*\dot{J} + \breve{J}*\breve{u} + \tfrac{1}{2}c\breve{u}*\breve{u} \\ - u*\hat{f} - u(t)\hat{j}(0) \tag{73}$$

Next, for the first variation of this mixed convolved action, we have the weak form

$$\delta I_C = \delta\dot{u}*m\,\dot{u} - \delta\dot{J}*a\,\dot{J} + \delta\breve{J}*\breve{u} + \delta\breve{u}*\breve{J} \\ + \delta\breve{u}*c\,\breve{u} - \delta u*\hat{f} - \delta u(t)\hat{j}(0) \tag{74}$$

After applying classical integration-by-parts (68b) to the first and second terms and fractional integration-by-parts (68a) to the third, fourth and fifth terms on the right-hand side of (74), we may write the following relation that corresponds to the stationarity of this mixed convolved action:

$$\begin{aligned}\delta I_C = &\ \delta u * [m\,\ddot{u} + c\,\dot{u} + \dot{J} - \hat{f}] + \delta J * [-a\,\ddot{J} + \dot{u}] \\ &+ \delta u(t)[m\,\dot{u}(0) + c\,u(0) + J(0) - \hat{j}(0)] + \delta u(0)[m\,\dot{u}(t)] \\ &+ \delta J(t)[-a\,\dot{J}(0) + u(0)] - \delta J(0)[-a\,\dot{J}(t)] = 0 \end{aligned} \quad (75)$$

For arbitrary variations $\delta u$ and $\delta J$, with the initial constraints

$$\delta u(0) = 0, \qquad \delta J(0) = 0, \quad (76a,b)$$

we recover from (75) all of the governing equations (70a), (70b), (71a) and (71b) in the strong form as the Euler-Lagrange equations associated with the mixed convolved action (73).

Thus, we have the *Principle of Mixed Convolved Action*. Of all the possible trajectories $\{u(\tau), J(\tau)\}$ of the dynamical single degree-of-freedom damped oscillator during the time interval $(0,t)$, the one that renders the action $I_C$, in (73), stationary corresponds to the solution of the equations of motion (70a) and compatibility (70b) throughout the time interval, while also satisfying the initial conditions defined by (71a) and (71b).

With this mixed convolved action approach, we are able to circumvent the Bauer (1931) Corollary. A single scalar functional $I_C$ is defined, which recovers all of the governing differential equations and initial conditions for our simple canonical dissipative system, while avoiding the introduction of any extraneous equations or constraints. Following the ideas of Gurtin (1964a,b) and especially Tonti (1973), this requires the replacement of the usual inner product with a convolution operator. This, in turn, implies that although the action $I_C$ is well-defined, there exists no Lagrangian or, in other words, the action is not simply the time integral of a state functional. In addition, following the proposal by Riewe (1996), the action $I_C$ incorporates fractional derivatives, which also represent convolutions. However, in order to model dissipation and recover the proper initial conditions of our dynamical system, we require the convolution of these convolutions in our action. Finally, following the approach of Sivaselvan and Reinhorn (2006), we pose the problem in mixed variables, involving both the displacement $u$ and the impulse of force in the spring $J$. Here, in order to recognize the full symmetry of the formulation, it is useful to view displacement as the impulse of the velocity of the mass.

In order to present the concept of the mixed convolved action in the simplest possible setting, we have kept the focus to this point in the paper on the single degree-of-freedom (SDOF) damped oscillator. However, the formalism has broad applicability throughout engineering science and mathematical physics. We begin the generalization in the next section by considering spatially discrete systems, involving the multi-degree-of-freedom case.

## 6. Principle of mixed convolved action for multiple degree of freedom oscillators

For a multi-degree-of-freedom (MDOF) system, let the state variables be the displacement vector from the equilibrium position $\mathbf{u}$ and the impulse of the internal elastic mechanical force vector $\mathbf{J}$. In integral form, these state variables are written

$$\mathbf{u} = \int_0^t \mathbf{v}\ dt \tag{77a}$$

$$\mathbf{J} = \int_0^t \mathbf{F}\ dt \tag{77b}$$

or in rate form

$$\dot{\mathbf{u}} = \mathbf{v} \tag{78a}$$

$$\dot{\mathbf{J}} = \mathbf{F} \tag{78b}$$

where $\mathbf{v}$ is the velocity vector, while $\mathbf{F}$ is the internal elastic force vector. Clearly, from (77a), one can think of the displacement as the impulse of velocity.

Generalizing the ideas from Section 5, we can now formulate a mixed convolved action for the MDOF system, as

$$I_C = \tfrac{1}{2}\dot{\mathbf{u}}^T * \mathbf{M}\dot{\mathbf{u}} - \tfrac{1}{2}\dot{\mathbf{J}}^T * \mathbf{A}\dot{\mathbf{J}} + \breve{\mathbf{J}}^T * \mathbf{B}^T \breve{\mathbf{u}} + \tfrac{1}{2}\breve{\mathbf{u}}^T * \mathbf{C}\breve{\mathbf{u}} - \mathbf{u}^T * \hat{\mathbf{f}} - \mathbf{u}^T(t)\hat{\mathbf{j}}(0) \tag{79}$$

where $\mathbf{M}$ is the nodal mass matrix; $\mathbf{A}$ is the element-based elastic flexibility matrix; $\mathbf{C}$ is the nodal damping matrix; and $\mathbf{B}$ is the equilibrium matrix. Here we assume that $\mathbf{M}$, $\mathbf{A}$, and $\mathbf{C}$ are all symmetric positive matrices and that $\mathbf{A}$ is block diagonal. The terms $\hat{\mathbf{f}}$ and $\hat{\mathbf{j}}(0)$ denote the specified applied nodal forces and the impulse of the applied forces at time zero, respectively. Again, we employ superposed dots to represent integer order time derivatives and a breve for left Riemann-Liouville fractional derivatives of one-half order.

The conditions for the stationarity of (79) can be found by setting the first variation to zero. Hence,

$$\delta I_C = \delta \dot{\mathbf{u}}^T * \mathbf{M}\dot{\mathbf{u}} - \delta \breve{\mathbf{J}}^T * \mathbf{A}\dot{\mathbf{J}} + \delta \breve{\mathbf{J}}^T * \mathbf{B}^T \breve{\mathbf{u}} + \delta \breve{\mathbf{u}}^T * \mathbf{B}\breve{\mathbf{J}} + \delta \breve{\mathbf{u}}^T * \mathbf{C}\breve{\mathbf{u}} \\ - \delta \mathbf{u}^T * \hat{\mathbf{f}} - \delta \mathbf{u}^T(t)\hat{\mathbf{j}}(0) = 0 \tag{80}$$

As in the SDOF case, we now use integral and fractional integration-by-parts to remove all of the temporal derivatives from the variations of $\delta \mathbf{u}$ and $\delta \mathbf{J}$. As a result, we have

$$\begin{aligned}\delta I_C &= \delta \mathbf{u}^T * \left[\mathbf{M}\ddot{\mathbf{u}} + \mathbf{C}\dot{\mathbf{u}} + \mathbf{B}\dot{\mathbf{J}} - \hat{\mathbf{f}}\right] + \delta \mathbf{J}^T * \left[-\mathbf{A}\ddot{\mathbf{J}} + \mathbf{B}^T \dot{\mathbf{u}}\right] \\ &+ \delta \mathbf{u}^T(t)\left[\mathbf{M}\dot{\mathbf{u}}(0) + \mathbf{C}\mathbf{u}(0) + \mathbf{B}\mathbf{J}(0) - \hat{\mathbf{j}}(0)\right] - \delta \mathbf{u}^T(0)\left[\mathbf{M}\dot{\mathbf{u}}(t)\right] \\ &+ \delta \mathbf{J}^T(t)\left[-\mathbf{A}\dot{\mathbf{J}}(0) + \mathbf{B}^T \mathbf{u}(0)\right] - \delta \mathbf{J}^T(0)\left[-\mathbf{A}\dot{\mathbf{J}}(t)\right] = 0\end{aligned} \tag{81}$$

The arbitrariness of the variations then leads to the following Euler-Lagrange equations governing this MDOF system

$$\mathbf{M}\ddot{\mathbf{u}} + \mathbf{C}\dot{\mathbf{u}} + \mathbf{B}\dot{\mathbf{J}} = \hat{\mathbf{f}} \tag{82a}$$

$$-\mathbf{A}\ddot{\mathbf{J}} + \mathbf{B}^T \dot{\mathbf{u}} = \mathbf{0} \tag{82b}$$

$$\mathbf{M}\dot{\mathbf{u}}(0) + \mathbf{C}\mathbf{u}(0) + \mathbf{B}\mathbf{J}(0) = \hat{\mathbf{j}}(0) \tag{83a}$$

$$-\mathbf{A}\dot{\mathbf{J}}(0) + \mathbf{B}^T \mathbf{u}(0) = \mathbf{0} \tag{83b}$$

Here, (82a) and (82b) represent the equations of motion and compatibility, respectively, and are valid throughout the time interval $(0,t)$. Equations (83a) and (83b) are the corresponding initial conditions. The variations are constrained only at the initial instant, such that $\delta \mathbf{u}(0) = \mathbf{0}$ and $\delta \mathbf{J}(0) = \mathbf{0}$.

For the *Principle of Mixed Convolved Action* of a dynamical multi-degree-of-freedom damped oscillatory system, we may state: Of all the possible trajectories $\{\mathbf{u}(\tau), \mathbf{J}(\tau)\}$ of the system during the time interval $(0,t)$, the one that renders the action $I_C$, in (79), stationary corresponds to the solution of the equations of motion (82a) and compatibility (82b) throughout the time interval, while also satisfying the initial conditions defined by (83a) and (83b).

## 7. Principle of mixed convolved action for elastic continua

As a third application, we consider the elastodynamic response of a continuum undergoing infinitesimal deformation. The classical continuum approach based upon Hamilton's principle is developed in Kirchhoff (1897) and later in Love (1927). Convolution-based continuum formulations appear first in Gurtin (1964b) and Tonti (1973). A broad range of these principles are collected and then extended by Oden and Reddy (1983) to Hellinger-Reissner type mixed principles. Here we apply the ideas from the previous sections to formulate a mixed variational principle that recovers the governing partial differential equations, along with all of the initial and boundary conditions for the elastodynamic problem.

For our mixed formulation of this problem, let the response be represented by the displacement $u_i$ and the impulse of the elastic stresses $J_{ij}$ (Sivaselvan and Reinhorn, 2006). Once again, for consistency, one can view displacement $u_i$ as the impulse of the velocity $v_i$. Thus, in integral form, we may write

$$u_i(t) = \int_0^t v_i(\tau) d\tau \tag{84a}$$

$$J_{ij}(t) = \int_0^t \sigma_{ij}(\tau) d\tau = \int_0^t C_{ijkl} \varepsilon_{kl}(\tau) d\tau \tag{84b}$$

or in rate form

$$\dot{u}_i = v_i \tag{85a}$$

$$\dot{J}_{ij} = \sigma_{ij} = C_{ijkl} \varepsilon_{kl} \tag{85b}$$

where $\sigma_{ij}$ and $\varepsilon_{ij}$ represent the stress and strain tensors, while $C_{ijkl}$ is the usual constitutive tensor for an anisotropic elastic medium. Beginning with (84), we use the standard indicial notation with repeated indices denoting summation over the spatial coordinate directions. In some cases, such as (84a) and (84b), we explicitly show the time dependency of the variables, whereas the spatial dependence on three-dimensional position $x$ is implicitly assumed.

Then, following the ideas developed in Sections 5 and 6 above, we may define a mixed convolved action as follows:

$$\begin{aligned} I_C = &\int_\Omega \left[ \tfrac{1}{2} \dot{u}_k * \rho \dot{u}_k - \tfrac{1}{2} \dot{J}_{ij} * A_{ijkl} \dot{J}_{kl} \right] d\Omega \\ &+ \int_\Omega \left[ \tfrac{1}{2} \left( \breve{J}_{ij} * B_{ijk} \breve{u}_k - \breve{u}_k * B_{ijk} \breve{J}_{ij} \right) \right] d\Omega \\ &- \int_\Omega \left[ u_k * \hat{f}_k + u_k(t) \hat{j}_k(0) \right] d\Omega \\ &- \int_{\Gamma_t} \tfrac{1}{2} \left[ u_k * \hat{t}_k + u_k(t) \hat{\tau}_k(0) \right] d\Gamma \\ &+ \int_{\Gamma_v} \tfrac{1}{2} \left[ \tau_k * \hat{v}_k + \tau_k(t) \hat{u}_k(0) \right] d\Gamma \end{aligned} \qquad (86)$$

where $\Omega$ denotes the spatial domain of the elastic body with bounding surface $\Gamma$. In (86), $\rho$ is the mass density and $A_{ijkl}$ is the elastic constitutive tensor inverse to $C_{ijkl}$, while $B_{ijk}$ is a third order tensor operator that extracts strain rates from the velocity field. Thus, we have

$$\dot{\varepsilon}_{ij} = B_{ijk} \dot{u}_k \qquad (87)$$

where more specifically

$$B_{ijk} = \frac{1}{2} \left( \delta_{ik} \delta_{jq} + \delta_{iq} \delta_{jk} \right) \frac{\partial}{\partial x_q} \qquad (88)$$

with $x_q$ representing spatial coordinates. The relation (87) may also be written in a more familiar manner as

$$\dot{\varepsilon}_{ij} = \tfrac{1}{2} \left( \dot{u}_{i,j} + \dot{u}_{j,i} \right) \qquad (89)$$

with the comma now symbolizing differentiation with respect to the spatial coordinates. Notice that we again use a superposed dot and breve to represent integer and left half order temporal derivatives, respectively. In addition, in (86), $\hat{f}_k$ and $\hat{t}_k$ represent the applied body force density and the applied surface tractions on a portion of the surface designated as $\Gamma_t$, while $\hat{j}_k(0)$ and $\hat{\tau}_k(0)$ are the impulses of $\hat{f}_k$ and $\hat{t}_k$ evaluated at time zero. In a similar way, $\hat{v}_k$ and $\hat{u}_k(0)$ represent the enforced surface velocities on $\Gamma_v$ and the corresponding impulses evaluated at the initial instant $t = 0$, while $\tau_k$ is the impulse of the traction $t_k$ on $\Gamma_v$. Here, we assume that the boundary conditions are defined, such that $\Gamma_v \cup \Gamma_t = \Gamma$ and $\Gamma_v \cap \Gamma_t = \emptyset$.

The first variation of the mixed convolved action becomes

$$\delta I_C = \int_\Omega \left[ \delta \dot{u}_k * \rho \dot{u}_k - \delta J_{ij} * A_{ijkl} J_{kl} \right] d\Omega$$
$$+ \int_\Omega \left[ \tfrac{1}{2} \left( \delta \breve{J}_{ij} * B_{ijk} \breve{u}_k - \delta \breve{u}_k * B_{ijk} \breve{J}_{ij} + B_{ijk} \delta \breve{u}_k * \breve{J}_{ij} - B_{ijk} \delta \breve{J}_{ij} * \breve{u}_k \right) \right] d\Omega$$
$$- \int_\Omega \left[ \delta u_k * \hat{f}_k + \delta u_k(t) \hat{j}_k(0) \right] d\Omega \quad (90)$$
$$- \int_{\Gamma_t} \tfrac{1}{2} \left[ \delta u_k * \hat{t}_k + \delta u_k(t) \hat{\tau}_k(0) \right] d\Gamma$$
$$+ \int_{\Gamma_v} \tfrac{1}{2} \left[ \delta \tau_k * \hat{v}_k + \delta \tau_k(t) \hat{u}_k(0) \right] d\Gamma$$

We need to perform temporal integration-by-parts on several terms in (90). However, we also require a spatial integration-by-parts operation on the terms involving $B_{ijk} \delta \breve{u}_k$ and $B_{ijk} \delta \breve{J}_{ij}$. For this development within the classical size-independent theory, we make use of the symmetry of stresses $\dot{J}_{kj}$ and the Cauchy definition of surface traction, where $t_k = \dot{\tau}_k = \dot{J}_{jk} n_j$. We also must recognize that similar relations hold for the left half-order time derivatives of these variables. The reformulation for $B_{ijk} \delta \breve{u}_k$ proceeds as follows:

$$\int_\Omega \left[ B_{ijk} \delta \breve{u}_k * \breve{J}_{ij} \right] d\Omega = \int_\Omega \left[ \tfrac{1}{2} \left( \delta \breve{u}_{i,j} + \delta \breve{u}_{j,i} \right) * \breve{J}_{ij} \right] d\Omega$$
$$= \int_\Omega \left[ \delta \breve{u}_{k,j} * \breve{J}_{kj} \right] d\Omega$$
$$= \int_\Omega \left[ \delta \breve{u}_k * \breve{J}_{kj} \right]_{,j} d\Omega - \int_\Omega \left[ \delta \breve{u}_k * \breve{J}_{kj,j} \right] d\Omega \quad (91a)$$
$$= \int_\Gamma \left[ \delta \breve{u}_k * \breve{J}_{jk} n_j \right] d\Gamma - \int_\Omega \left[ \delta \breve{u}_k * \breve{J}_{jk,j} \right] d\Omega$$
$$= \int_\Gamma \left[ \delta \breve{u}_k * \breve{\tau}_k \right] d\Gamma - \int_\Omega \left[ \delta \breve{u}_k * \breve{J}_{jk,j} \right] d\Omega$$
$$= \int_\Gamma \left[ \delta \breve{u}_k * \breve{\tau}_k \right] d\Gamma - \int_\Omega \left[ \delta \breve{u}_k * B_{ijk} \breve{J}_{ij} \right] d\Omega$$

Similarly, for $B_{ijk} \delta \breve{J}_{ij}$, we have

$$\int_\Omega \left[ B_{ijk} \delta \breve{J}_{ij} * \breve{u}_k \right] d\Omega = \int_\Gamma \left[ \delta \breve{\tau}_k * \breve{u}_k \right] d\Gamma - \int_\Omega \left[ \delta \breve{J}_{ij} * B_{ijk} \breve{u}_k \right] d\Omega \quad (91b)$$

After substituting (91a) and (91b) into (90), we have

$$\delta I_C = \int_\Omega \left[ \delta \dot{u}_k * \rho \dot{u}_k - \delta \breve{J}_{ij} * A_{ijkl} \breve{J}_{kl} \right] d\Omega$$
$$+ \int_\Omega \left[ \delta \breve{J}_{ij} * B_{ijk} \breve{u}_k - \delta \breve{u}_k * B_{ijk} \breve{J}_{ij} \right] d\Omega$$
$$+ \int_\Gamma \tfrac{1}{2} \left[ \delta \breve{u}_k * \breve{\tau}_k - \delta \breve{\tau}_k * \breve{u}_k \right] d\Gamma$$
$$- \int_\Omega \left[ \delta u_k * \hat{f}_k + \delta u_k(t) \hat{j}_k(0) \right] d\Omega \quad (92)$$
$$- \int_{\Gamma_t} \tfrac{1}{2} \left[ \delta u_k * \hat{t}_k + \delta u_k(t) \hat{\tau}_k(0) \right] d\Gamma$$
$$+ \int_{\Gamma_v} \tfrac{1}{2} \left[ \delta \tau_k * \hat{v}_k + \delta \tau_k(t) \hat{u}_k(0) \right] d\Gamma$$

After performing all of the temporal integration-by-parts operations, the stationarity of the mixed convolved action may be written

$$\delta I_C = \int_\Omega \delta u_k * \left[ \rho \ddot{u}_k - B_{ijk} \dot{J}_{ij} - \hat{f}_k \right] d\Omega + \int_\Omega \delta J_{ij} * \left[ -A_{ijkl} \ddot{J}_{kl} + B_{ijk} \dot{u}_k \right] d\Omega$$
$$+ \int_\Omega \delta u_k(t) \left[ \rho \dot{u}_k(0) - B_{ijk} J_{ij}(0) - \hat{j}_k(0) \right] d\Omega - \int_\Omega \delta u_k(0) \left[ \rho \dot{u}_k(t) \right] d\Omega$$
$$+ \int_\Omega \delta J_{ij}(t) \left[ -A_{ijkl} \dot{J}_{kl}(0) + B_{ijk} u_k(0) \right] d\Omega - \int_\Omega \delta J_{ij}(0) \left[ -A_{ijkl} \dot{J}_{kl}(t) \right] d\Omega$$
$$+ \int_{\Gamma_t} \tfrac{1}{2} \left[ \delta u_k * t_k - \delta u_k * \hat{t}_k \right] d\Gamma + \int_{\Gamma_t} \tfrac{1}{2} \left[ \delta u_k(t) \tau_k(0) - \delta u_k(t) \hat{\tau}_k(0) \right] d\Gamma \quad (93)$$
$$+ \int_{\Gamma_v} \tfrac{1}{2} \left[ \delta u_k * t_k \right] d\Gamma + \int_{\Gamma_v} \tfrac{1}{2} \left[ \delta u_k(t) \tau_k(0) \right] d\Gamma$$
$$- \int_{\Gamma_v} \tfrac{1}{2} \left[ \delta \tau_k * v_k - \delta \tau_k * \hat{v}_k \right] d\Gamma - \int_{\Gamma_v} \tfrac{1}{2} \left[ \delta \tau_k(t) u_k(0) - \delta \tau_k(t) \hat{u}_k(0) \right] d\Gamma$$
$$- \int_{\Gamma_t} \tfrac{1}{2} \left[ \delta \tau_k * v_k \right] d\Gamma - \int_{\Gamma_t} \tfrac{1}{2} \left[ \delta \tau_k(t) u_k(0) \right] d\Gamma = 0$$

From (93), we have as the Euler-Lagrange equations:

$$\rho \ddot{u}_k - B_{ijk} \dot{J}_{ij} = \hat{f}_k \qquad x \in \Omega,\ \tau \in (0,t) \qquad (94a)$$
$$-A_{ijkl} \ddot{J}_{kl} + B_{ijk} \dot{u}_k = 0 \qquad x \in \Omega,\ \tau \in (0,t) \qquad (94b)$$
$$t_k = \hat{t}_k \qquad x \in \Gamma_t,\ \tau \in (0,t) \qquad (95a)$$
$$v_k = \hat{v}_k \qquad x \in \Gamma_v,\ \tau \in (0,t) \qquad (95b)$$
$$\rho \dot{u}_k(0) - B_{ijk} J_{ij}(0) = \hat{j}_k(0) \qquad x \in \Omega \qquad (96a)$$
$$-A_{ijkl} \dot{J}_{kl}(0) + B_{ijk} u_k(0) = 0 \qquad x \in \Omega \qquad (96b)$$
$$\tau_k(0) = \hat{\tau}_k(0) \qquad x \in \Gamma_t \qquad (97a)$$

$$u_k(0) = \hat{u}_k(0) \qquad x \in \Gamma_v \qquad (97b)$$

with the variations defined, such that

$$\delta \tau_k = 0 \qquad x \in \Gamma_t, \; \tau \in (0,t) \qquad (98a)$$

$$\delta u_k = 0 \qquad x \in \Gamma_v, \; \tau \in (0,t) \qquad (98b)$$

$$\delta u_k(0) = 0 \qquad x \in \Omega \qquad (99a)$$

$$\delta J_{ij}(0) = 0 \qquad x \in \Omega \qquad (99b)$$

$$\delta \tau_k(t) = 0 \qquad x \in \Gamma_t \qquad (100a)$$

$$\delta u_k(0) = 0 \qquad x \in \Gamma_v \qquad (100b)$$

Consequently, we may write a *Principle of Mixed Convolved Action* for an elastodynamic continuum undergoing infinitesimal deformation as follows: Of all the possible trajectories $\{u_k(\tau), J_{ij}(\tau)\}$ of the system during the time interval $(0,t)$, the one that renders the action $I_C$, in (86), stationary corresponds to the solution of the initial/boundary value problem. Thus, the stationary trajectory satisfies the equations of motion (94a) and compatibility (94b) in the domain $\Omega$, along with the traction (95a) and velocity (95b) boundary conditions, throughout the time interval, while also satisfying the initial conditions defined by (96a) and (96b) in $\Omega$ and (97a) and (97b) on the appropriate portions of the bounding surface. Furthermore, the possible trajectories under consideration during the variational process are constrained only by their need to satisfy the specified initial and boundary conditions of the problem in the form of (98a,b), (99a,b) and (100a,b).

## 8. Conclusions

The concept of mixed convolved action is advanced as a means to formulate stationary variational principles for a broad range of dynamical systems. In particular, this resolves the long-standing difficulty of defining a scalar functional for dissipative systems. Furthermore, it provides a direct connection between the strong form of the dynamical initial/boundary value problem and the weak statement by restricting variations in a manner consistent with the specification of the initial and boundary conditions.

We consider here the development of a mixed convolved action for the classical single degree-of-freedom damped oscillator, multi-degree-of-freedom structural dynamics and an elastodynamic continuum. All of these illustrate the inner symmetries and elegance of the approach. Clearly, however, the concept of mixed convolved action is quite general and can be applied readily throughout mathematical physics and engineering. The transformation of these formulations to the frequency domain also could provide further physical insight. In addition, we anticipate that the weak forms developed here, and those to come in related fields, will provide an interesting foundation for the development of novel computational methods.

## References

Arthurs, A. M., Jones, M. E., 1976. On variational principles for linear initial value problems, *Journal of Mathematical Analysis and Applications*. 54, 840–845.

Bateman, H., 1931. On dissipative systems and related variational pronciples, *Physical Review*. 38, 815-819.

Bauer, P. S., 1931. Dissipative dynamical systems. I., *Proceedings of the National Academy of Sciences*. 17, 311-314.

Cresson, J., 2007. Fractional embedding of differential operators and Lagrangian systems, *Journal of Mathematical Physics*. 48, 033504.

Davis, D. R., 1928. The inverse problem of the calculus of variations in higher space, *Transactions of the American Mathematical Society*. 30(4), 710-736.

Gurtin, M. E., 1964a. Variational principles for linear initial-value problems, *Quarterly of Applied Mathematics*. 22(3), 252-256.

Gurtin, M. E., 1964b. Variational principles for linear elastodynamics, *Archive for Rational Mechanics and Analysis*. 16(1), 34-50.

Hardy, G. H., Littlewood, J. E., 1928. Some properties of fractional integrals, I, *Mathematische Zeitschrift*. 27, 565-606.

José, J. V., Saletan, E. J., 1998. *Classical dynamics: A contemporary approach*. Cambridge University Press, Cambridge, UK.

Kirchhoff, G., 1897. *Vorlesungen über mechanic*. B. G. Teubner, Leipzig.

Lanczos, C., 1949. *The variational principles of mechanics*. University of Toronto Press, Toronto.

Love, A. E. H., 1927. *A treatise on the mathematical theory of elasticity*. Cambridge University Press, Cambridge.

Love, E. R., Young, L. C., 1938. On fractional integration by parts. *Proceedings of the London Mathematical Society*, Series 2. 44, 1-35.

Morse, M., 1930. A generalization of the Sturm, separation and comparison theorems in n-space. *Mathematische Annalen*. 103, 52-69.

Morse, P. M., Feshbach, H., 1953. *Methods of theoretical physics*. McGraw-Hill, New York.

Oden, J. T., Reddy, J. N., 1983. *Variational methods in theoretical mechanics*. Springer-Verlag, Berlin.

Oldham, K. B., Spanier, J., 1974. *The fractional calculus*. Academic Press, New York.

Rayleigh, J. W. S., 1877. *The theory of sound*. I & II. Second edition, reprint in 1945. Dover Publications, New York.

Riewe, F., 1996. Nonconservative Lagrangian and Hamiltonian mechanics. *Physical Review E*. 53, 1890-1899.

Riewe, F., 1997. Mechanics with fractional derivatives. *Physical Review E*. 55, 3581-3592.

Ross, B., 1977. Fractional calculus. *Mathematics Magazine*. 50(3), 115-122.

Samko, S. G., Kilbas, A. A., Marichev, O. I., 1993. *Fractional integrals and derivatives*. Gordon and Breach Science Publishers, Switzerland.

Sivaselvan, M. V., Reinhorn, A. M., 2006. Lagrangian approach to structural collapse simulation. *Journal of Engineering Mechanics*, ASCE. 132, 795-805.

Sivaselvan, M. V., Lavan, O., Dargush, G. F., *et al.*, 2009. Numerical collapse simulation of large-scale structural systems using an optimization-based algorithm. *Earthquake Engineering and Structural Dynamics*. 38, 655-677.

Tonti, E., 1973. On the variational formulation for linear initial value problems. *Annali di Matematica pura ed applicata*. XCV, 331-360.

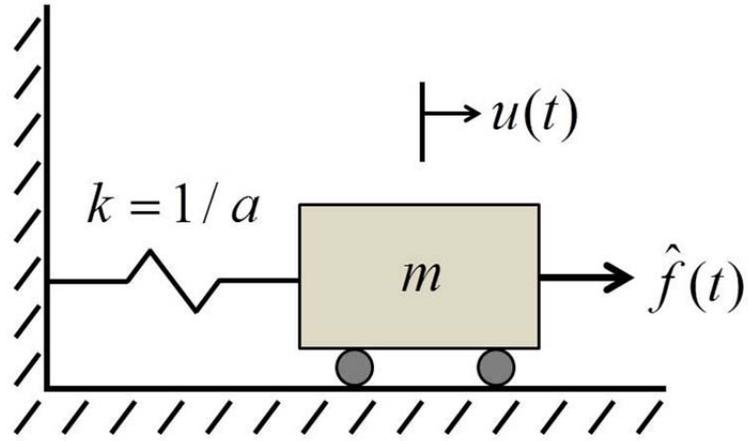

**Fig. 1.** Single-degree-of-freedom oscillator

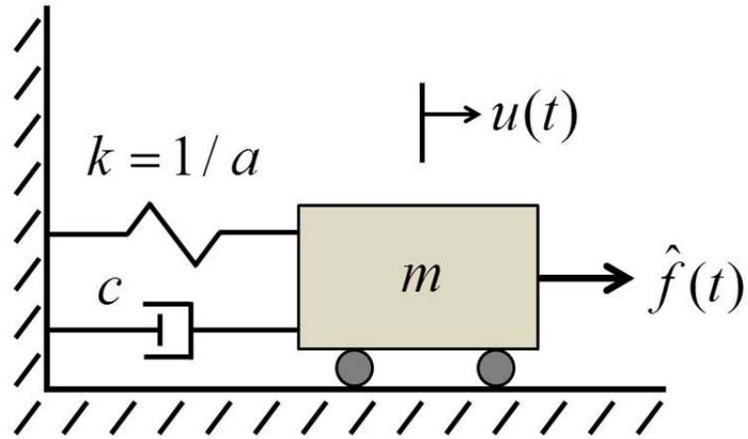

**Fig. 2.** Single-degree-of-freedom damped oscillator